\documentclass[final,5p,times,twocolumn]{elsarticle}

\usepackage{amssymb}

\journal{Physics Letters B}

\begin{document}

\begin{frontmatter}

\title{Search for $^{6}_{\Lambda}$H hypernucleus by the $^6$Li($\pi^-,K^+$) reaction at $p_{\pi^-}$ = 1.2 GeV/$c$}

\author[kyoto,jaea]{H.~Sugimura\corref{cor1}}
\ead{sugimura@scphys.kyoto-u.ac.jp}
\cortext[cor1]{Corresponding author. Tel:+81 29 2825457}
\author[torino,infn]{M.~Agnello}
\author[pnu]{J.K.~Ahn}
\author[rcnp]{S.~Ajimura}
\author[tohoku]{Y.~Akazawa}
\author[kyoto]{N.~Amano}
\author[kek]{K.~Aoki}
\author[snu]{H.C.~Bhang}
\author[tohoku]{N.~Chiga}
\author[osaka]{M.~Endo}
\author[jinr]{P.~Evtoukhovitch}
\author[infn]{A.~Feliciello}
\author[kyoto]{H.~Fujioka}
\author[oec]{T.~Fukuda}
\author[jaea]{S.~Hasegawa}
\author[osaka]{S.~Hayakawa}
\author[tohoku]{R.~Honda}
\author[tohoku]{K.~Hosomi}
\author[jaea]{S.H.~Hwang}
\author[kyoto,jaea]{Y.~Ichikawa}
\author[kek]{Y.~Igarashi}
\author[jaea]{K.~Imai}
\author[osaka]{N.~Ishibashi}
\author[kek]{R.~Iwasaki}
\author[snu]{C.W.~Joo}
\author[snu,jaea]{R.~Kiuchi}
\author[pnu]{J.K.~Lee}
\author[snu]{J.Y.~Lee}
\author[osaka]{K.~Matsuda}
\author[tohoku]{Y.~Matsumoto}
\author[osaka]{K.~Matsuoka}
\author[tohoku]{K.~Miwa}
\author[oec]{Y.~Mizoi}
\author[rcnp]{M.~Moritsu}
\author[kyoto]{T.~Nagae}
\author[jaea]{S.~Nagamiya}
\author[osaka]{M.~Nakagawa}
\author[kyoto]{M.~Naruki}
\author[rcnp]{H.~Noumi}
\author[osaka]{R.~Ota}
\author[barc]{B.J.~Roy}
\author[jaea]{P.K.~Saha}
\author[osaka]{A.~Sakaguchi}
\author[jaea]{H.~Sako}
\author[vmi]{C.~Samanta}
\author[jinr]{V.~Samoilov}
\author[tohoku]{Y.~Sasaki}
\author[jaea]{S.~Sato}
\author[kek]{M.~Sekimoto}
\author[oec]{Y.~Shimizu}
\author[tohoku]{T.~Shiozaki}
\author[rcnp]{K.~Shirotori}
\author[osaka]{T.~Soyama}
\author[kek]{T.~Takahashi}
\author[riken]{T.N.~Takahashi}
\author[tohoku]{H.~Tamura}
\author[tohoku]{K.~Tanabe}
\author[osaka]{T.~Tanaka}
\author[snu]{K.~Tanida}
\author[rcnp]{A.O.~Tokiyasu}
\author[jinr]{Z.~Tsamalaidze}
\author[tohoku]{M.~Ukai}
\author[tohoku]{T.O.~Yamamoto}
\author[tohoku]{Y.~Yamamoto}
\author[snu]{S.B.~Yang}
\author[osaka]{K.~Yoshida}

\author[]{(J-PARC E10 Collaboration)}

\address[kyoto]{Department of Physics, Kyoto University, Kyoto 606-8502, Japan}
\address[jaea]{Japan Atomic Energy Agency(JAEA), Tokai, Ibaraki 319-1195, Japan}
\address[torino]{Dipartimento di Scienza Applicata e Tecnologia, Politecnico di Torino, I-10129, Torino, Italy}
\address[infn]{INFN, Istituto Nazionale di Fisica Nucleare, Sez. di Torino I-10125, Torino, Italy}
\address[pnu]{Department of Physics, Pusan National University, Busan 609-735, Republic of Korea}
\address[rcnp]{Research Center for Nuclear Physics (RCNP), 10-1 Mihogaoka, Ibaraki, Osaka 567-0047, Japan}
\address[tohoku]{Department of Physics, Tohoku University, Sendai 980-8578, Japan}
\address[kek]{High Energy Accelerator Research Organization (KEK), Tsukuba 305-0801, Japan}
\address[snu]{Department of Physics and Astronomy, Seoul National University, Seoul 151-747, Republic of Korea}
\address[osaka]{Department of Physics, Osaka University, Toyonaka, Osaka 560-0043, Japan}
\address[jinr]{Joint Institute for Nuclear Research, Dubna, Moscow Region 141980, Russia}
\address[oec]{Department of Engineering Science, Osaka Electro-Communication University, Neyagawa, Osaka 572-8530, Japan}
\address[barc]{Nuclear Physics Division, Bhabha Atomic Research Center (BARC), Trombay, Mumbai 400 085, India}
\address[vmi]{Department of Physics and Astronomy, Virginia Military Institute, Lexington, VA 24450, USA}
\address[riken]{RIKEN, 2-1 Hirosawa, Wako, Saitama 351-0198, Japan}

\begin{abstract}
We have carried out an experiment to search for a neutron-rich hypernucleus,
$^6_{\Lambda}$H, by the $^6$Li($\pi^-,K^+$) reaction at $p_{\pi^-}$ =1.2 GeV/$c$.
The obtained missing-mass spectrum with an estimated energy resolution of
3.2 MeV (FWHM) showed no peak structure corresponding
to the $^6_{\Lambda}$H hypernucleus neither below nor above the $^4_{\Lambda}$H$+2n$
particle decay threshold.
An upper limit of the production cross section for the bound $^6_{\Lambda}$H
hypernucleus was estimated to be 1.2 nb/sr at 90\% confidence level.
\end{abstract}


\end{frontmatter}

\section{Introduction}
\label{intro}

Since the first discovery of a cosmic-ray induced hyperfragment formation event
in emulsion \cite{Danysz}, various $\Lambda$-hypernuclei have been observed. 
Especially, extensive studies have been made by missing-mass spectroscopy
with magnetic spectrometers by using the ($K^-,\pi^-$) and ($\pi^+,K^+$)
reactions \cite{bona,may,pile,hotchi}, and the properties of the $\Lambda$N
interaction have been extracted from the level structures of
the $\Lambda$-hypernuclei.
Furthermore, details of the spin-dependent interactions have been
studied by $\gamma$-ray spectroscopy with germanium and
NaI detectors \cite{tamura,ajimura,ukai}.

When we embed a $\Lambda$ hyperon in a nucleus, there are several
interesting features theoretically expected for some specific hypernuclei.
One example is the glue-like role of a $\Lambda$ hyperon in the hypernuclei.
Unstable nuclei such as $^8$Be become stable against the particle
decays by adding a $\Lambda$ hyperon.
The glue-like role gives a critical contribution to
the binding especially around the proton- and neutron-drip lines,
and may extend the boundary of stability of nuclei.
Another example is a large effect of the
$\Lambda$N-$\Sigma$N mixing which was first discussed by Gibson et al. for
$s$-shell $\Lambda$-hypernuclei \cite{gibson}.
The idea has been extended to the coherent $\Lambda$N-$\Sigma$N mixing
by Akaishi et al. to understand the binding energies of
$s$-shell $\Lambda$-hypernuclei systematically \cite{Akaishi1}.
A non-zero isospin of the core nucleus is essential for
the large mixing because the core nucleus is a buffer of the isospin to
compensate the isospin difference between $\Lambda$ and $\Sigma$.
The studies on the neutron-rich $\Lambda$-hypernuclei with
a large isospin are important to understand the
properties of the $\Lambda$N-$\Sigma$N mixing effect.

The highly neutron-rich $\Lambda$-hypernucleus $^6_{\Lambda}$H
was first discussed by Dalitz and Levi-Setti \cite{Dalitz1}.
They predicted the $\Lambda$-binding energy of the ground state to be 4.2 MeV
based on the knowledge of the core nucleus $^5$H at that time.
Later on, \mbox{Korsheninnikov} et al. reported an observation of the
$^5$H ground state as a resonant state unbound by 1.7 MeV 
with respect to the $t+2n$ threshold \cite{dubna5H}.
Including the new experimental information,
there were extensive theoretical discussions on the structure
of $^6_{\Lambda}$H.
Akaishi et al. suggested a considerably large binding energy
of 5.8 MeV for the 0$^+$ ground state due to rather large contribution
of 1.4 MeV from the coherent $\Lambda$N-$\Sigma$N mixing \cite{Akaishi2}.
Gal and Millener predicted a binding energy of $(3.83 \pm 0.08 \pm 0.22)$ MeV
by a shell-model calculation \cite{gal}.
Hiyama et al. performed $t + n + n + \Lambda$ four-body cluster-model
calculation, which reproduced the properties of $^5$H as well,
and obtained a binding energy of 2.47 MeV \cite{hiyama}.
These theoretically estimated binding
energies were distributed from bound to unbound regions with respect to
the $^4_{\Lambda}$H$+2n$ particle decay threshold, which corresponded to
a $\Lambda$-binding energy of 3.74 MeV, and were quite sensitive
to the $\Lambda$N interaction and the properties of the core nucleus $^5$H.

The double charge-exchange (DCX) reactions, such as the ($K^-,\pi^+$)
and the ($\pi^-,K^+$) reactions, are promising spectroscopic tools to access
the $\Lambda$-hypernuclei close to the neutron drip-line \cite{Majling}.
The first study of the neutron-rich $\Lambda$-hypernuclei
with the DCX reaction was performed at the KEK proton synchrotron facility
by using the $(K^-_{stopped},\pi^+)$ reaction, and upper limits of 
the production cross sections for
several neutron-rich $\Lambda$-hypernuclei were provided \cite{kubota}.
Another DCX reaction, the $^{10}$B($\pi^-,K^+$) reaction at 1.2 GeV/$c$,
was measured in the \mbox{KEK-E521} experiment \cite{saha}.
Since there was no physical background in the ($\pi^-,K^+$) reaction,
the $^{10}_{~\Lambda}$Li production events were clearly observed in
the $\Lambda$ bound region.
The production cross section was reported
$(11.3 \pm 1.9)$ nb/sr which was roughly 10$^{-3}$ of that of
the non charge-exchange ($\pi^+,K^+$) reaction.
%Harada et al. calculated the reaction mechanism of the double charge-exchange
%reaction by using KEK-E521 experimental data~\cite{harada},
%but there is no theoretical calculation on the $^6$Li$(\pi^-,K^+$)$^6_{\Lambda}$H.
Possible reaction mechanisms of the ($\pi^-,K^+$) reaction were
discussed by Harada et al. from a theoretical point of view~\cite{harada}.
The FINUDA collaboration also measured the ($K^-_{stopped},\pi^+$) reaction
\cite{finuda1} and recently reported
three candidate events of $^6_{\Lambda}$H by the simultaneous measurement of both
the production and the mesonic weak decay processes
of $^6_{\Lambda}$H \cite{finuda2}.
The FINUDA collaboration reported a binding energy for $^6_\Lambda$H
close to the Dalitz and Levi-Setti calculation.
However, because of the small number of the candidate events,
the FINUDA observation should be confirmed with higher statistics.

\section{J-PARC E10 experiment}

\begin{figure}[htb]
  \centering
  \includegraphics[width=60mm]{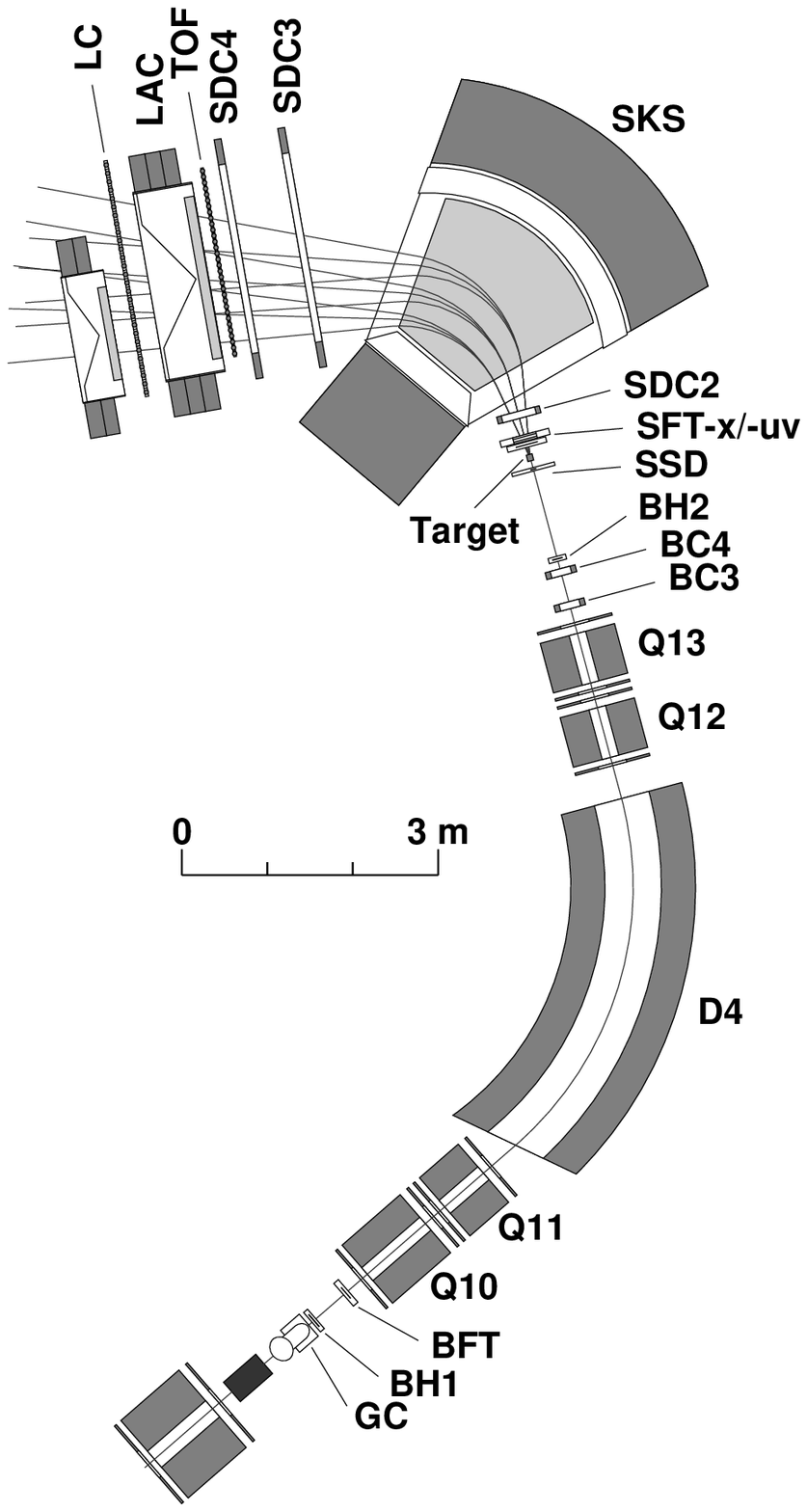}
  \caption{Schematic view of the K1.8 beam line spectrometer (from GC to BH2), the target area (SSD and Target) and the SKS spectrometer (from SFT to LC). See the text for more details.}
  \label{schema}
\end{figure}

The J-PARC E10 experiment was proposed to produce the neutron-rich
$\Lambda$-hypernucleus $^6_{\Lambda}$H by using the $^6$Li($\pi^-,K^+$) reaction
at 1.2 GeV/$c$ and to study its structure.
The experiment is performed at the K1.8 beam line of J-PARC
Hadron Experimental Facility.
The K1.8 beam line spectrometer \cite{takahasi} and the Superconducting
Kaon Spectrometer (SKS) \cite{takahasi,fukuda} are used after modifications
to cope with the high beam intensity,
more than 10$^7$ pions per spill (2s duration) occurring every 6s,
which is necessary to override the tiny production cross section.
Figure \ref{schema} shows the K1.8 beam line and the SKS spectrometers.

%%% K1.8 beam line spectrometer
The K1.8 beam line spectrometer consists of
a gas {\v C}erenkov counter (GC),
a scintillating fiber tracker (BFT), QQDQQ magnets,
two drift chambers (BC3 and BC4)
and a timing plastic scintillation hodoscope (BH2).
% BFT
BFT is a  scintillating fiber tracker with a xx$^\prime$ structure
and is made of scintillating fibers with 1 mm diameter.
The light signals from the scintillating fibers are read out
by MPPCs (Multi Pixel Photon Counters) and the EASIROC system \cite{honda}.
The time resolution is 0.86 ns (rms) which is good enough to
reduce accidental hits due to the high rate beams.
% BC3,4
BC3 and BC4 are 3 mm wire pitch drift chambers.
% BH2
BH2 is a timing plastic scintillator placed at
the exit of the K1.8 beam line spectrometer.
BH2 defines the start timing of the readout and the DAQ systems 
and counts the number of incoming pions.
The time resolution of BH2 is 90 ps (rms).
% Momentum
The pion beam momentum is reconstructed to an accuracy of $3.3 \times 10^{-4}$
(FWHM) by using a 3rd-order transfer matrix and the hit position
information from BFT, BC3, and BC4.

%%% around the target 
An enriched $^6$Li target (95.54\%) of 3.5 g/cm$^2$ in thickness,
70 mm in width and 40 mm in height is used.
%$70^W {\times} 40^H$(mm$^2$) in size is used. 
The typical beam profile, 56 mm in horizontal and 28 mm in vertical (FWTM),
is fully covered with the target.
To reconstruct the reaction vertex point precisely and to find the right
combination of upstream and downstream tracks for multi-track
events, silicon strip detectors (SSD), with a 80 $\mu$m strip pitch,
are installed just upstream the target.

%%% SKS spectrometer
The SKS spectrometer consists of
a scintillating fiber tracker (SFT), three
drift chambers (SDC2, SDC3 and SDC4), a superconducting magnet,
a timing plastic scintillator hodoscope (TOF)
and two threshold-type {\v C}erenkov counters (LAC and LC).
The SKS magnetic field is 2.16 T and 
the momentum acceptance of the SKS spectrometer ranges from 0.7 to 1.4 GeV/$c$.
The kaon momentum from the $^6$Li($\pi^-,K^+$)$^6_{~\Lambda}$H reaction is around
0.9 GeV/$c$, while the angular acceptance of the SKS spectrometer is roughly
100 msr at that momentum.
% SFT
SFT is a scintillating
fiber tracker with a xx$^\prime$vu structure.
The x and x$^\prime$ planes have a similar structure of BFT.
The v and u planes tilted by 45 degrees are made of
scintillating fibers with 0.5 mm diameter.
The time resolutions are
0.87 ns and 1.17 ns (rms) for the xx$^\prime$ and
uv planes, respectively.
% SDC2
SDC2 is a 5 mm wire pitch drift chamber.
SFT and SDC2 are placed at the entrance of the SKS magnet.
% SDC34
SDC3 and SDC4 are 20 mm wire pitch drift chambers 
placed at the exit of the SKS magnet.
% TOF
TOF is a timing plastic scintillator and the 
timing resolution is 85 ps (rms).
% LAC and LC
LAC and LC are threshold-type {\v C}erenkov counters
with a $n$=1.05 silica aerogel and a $n$=1.49 acrylic radiators, respectively.
Kaons do not exceed the {\v C}erenkov threshold of LAC and exceed the threshold
of LC for the momentum range covered by SKS.
The scattered particle momentum is reconstructed to an accuracy of
0.1$\%$ (FWHM) by using Runge-Kutta integration method with
the hit position information of SFT, SDC2, SDC3 and SDC4.

% Trigger
The on-line trigger for the measurements  of the
($\pi^\pm,K^+$) reactions consists of the 
1st level trigger defined as 
BH2 ${\times}$ TOF ${\times}$ $\overline{{\rm LAC}}$ ${\times}$ LC and
the 2nd level trigger to reduce proton backgrounds by using the BH2--TOF
 time-of-flight information.
% DAQ
The data are taken by a
network based DAQ system, HDDAQ \cite{igarashi},
which integrates several kinds of conventional DAQ subsystems.

\section{Results}

\begin{figure}[hbt]
  \centering
\includegraphics[width=85mm]{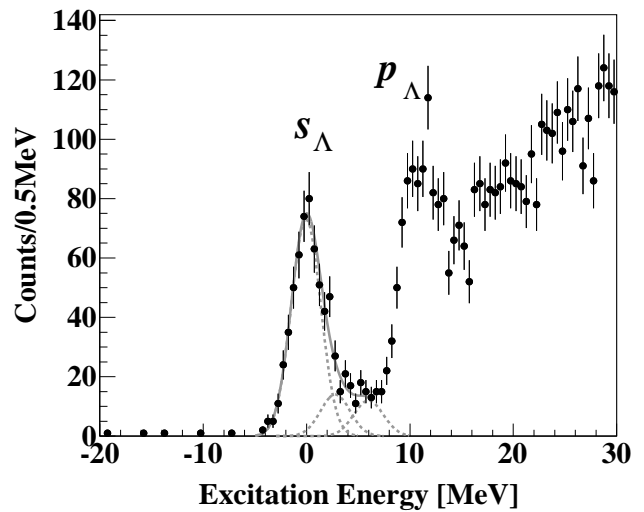}
\caption{Excitation energy spectrum of the $^{12}$C$(\pi^+,K^+)^{12}_{~\Lambda}$C reaction at the beam momentum of 1.2 GeV/$c$. The ground ($s_{\Lambda}$) and excited ($p_{\Lambda}$) states are clearly observed. The missing-mass resolution is estimated by fitting the ground and known excited states. The dashed curves show the best fit Gaussian functions for these states and the solid curve is the sum.
}
\label{12CL}
\end{figure}

Table \ref{run_summary} shows a summary of runs of the J-PARC E10 experiment.
\begin{table*}[hbt]
\caption{Summary of runs.
   Target material and thickness, and beam momentum and intensity are listed.}
\centering
\label{run_summary}       % Give a unique label
\begin{tabular}{cccccc}
\hline
 & \multicolumn{2}{c}{target} & & \multicolumn{2}{c}{beam} \\
\cline{2-3}\cline{5-6}
run & material & thickness (g/cm$^2$) & & momentum (GeV/$c$) & intensity (pion/spill) \\
\hline
$^6$Li$(\pi^-,K^+)$ & $^6$Li (95.54\% enriched) & 3.5  & & 1.2 & $1.2$--$1.4 \times 10^7$ \\
$^{12}$C$(\pi^+,K^+)$ & graphite & 3.6 & & 1.2 & $4.1 \times 10^6$ \\
$p(\pi^-,K^+)\Sigma^-$ & (CH$_2$)$_n$ & 3.4 & & 1.377 & $1.3 \times 10^7$ \\
$p(\pi^+,K^+)\Sigma^+$ & (CH$_2$)$_n$ & 3.4 & & 1.377 & $3.5 \times 10^6$ \\
beam-through & none & & & 0.8, 0.9, 1.0, 1.2 & $\sim 10^4$ \\
\hline
\end{tabular}
% Or use
%\vspace*{5cm}  % with the correct table height
\end{table*}
%%% Production run
% 12CL
Before the measurement of the DCX reaction on the $^6$Li target, we 
measured the $^{12}$C($\pi^+,K^+$)$^{12}_{~\Lambda}$C reaction at the beam momentum 
of 1.2 GeV/$c$ to evaluate the missing-mass resolution in a kinematical
condition close to that of the $^6$Li$(\pi^-,K^+)^6_\Lambda$H reaction.
A graphite target of 3.6 g/cm$^2$ in thickness was used,  and the beam
intensity was $4.1 \times 10^6$ pion/spill.
Figure \ref{12CL} shows the excitation energy spectrum of the 
$^{12}_{~\Lambda}$C hypernucleus obtained from the measurement.
The ground ($s_{\Lambda}$) and excited  ($p_{\Lambda}$) states
of the $^{12}_{~\Lambda}$C hypernucleus are clearly observed.
The ground state region is
fitted with three Gaussian functions corresponding to
the ground and to known excited states at 2.833 and 6.050 MeV \cite{ma,hosomi}.
A same width of the Gaussian function is used for the three states.
The missing-mass resolution is estimated to be 3.2 MeV (FWHM).

% Beam Through
We performed the pion beam-through runs at four momentum settings,
0.8, 0.9, 1.0 and 1.2 GeV/$c$,
without the target to evaluate the momentum difference
between the K1.8 beam line and the SKS spectrometers.
% Sigma +-
We also measured the $p$($\pi^-,K^+$)$\Sigma^-$ and
the $p$($\pi^+,K^+$)$\Sigma^+$ reactions at 1.377 GeV/$c$ to calibrate
the beam momentum with a (CH$_2$)$_n$ target of 3.4 g/cm$^2$ in thickness.
The pion beam momentum 1.377 GeV/$c$ was selected so that the produced
$K^+$ momentum in the $p$($\pi^\pm,K^+$)$\Sigma^\pm$ reaction
coincides with that in the $^6$Li($\pi^-,K^+$)$^6_{\Lambda}$H reaction
at 1.2 GeV/$c$.
The beam intensities in the $\Sigma^-$ and $\Sigma^+$ production
runs were $1.3 \times 10^7$ and 
$3.5 \times 10^6$ pion/spill, respectively.
%% Correction of momentum
The momenta of particles measured by the K1.8 beam line spectrometer 
are corrected with a 1st-order polynomial function
by considering the results of the beam-through
run at 0.9 GeV/$c$ and the $\Sigma^+$ production reaction.
Furthermore, the effect of the beam polarity change from $\pi^+$ to $\pi^-$
to the missing-mass
is evaluated by the $\Sigma^-$ production reaction.
The momenta of scattered particles
measured by the SKS spectrometer are not corrected
because the SKS magnetic field is fixed in all runs.
In the case of the 1.2 GeV/$c$ beam momentum used in the
$^6$Li($\pi^-,K^+$) reaction measurement,
the amount of the momentum correction is $-1.2$ MeV/$c$.
The systematic uncertainty of the beam momentum is estimated from the
beam-through runs at 0.8, 1.0 and 1.2 GeV/$c$ after the momentum correction.
The momentum differences after the correction are -1.46, +1.62, and
-0.79 MeV/$c$ at 0.8, 1.0, and 1.2 GeV/$c$, respectively.
From these values, the systematic uncertainty is estimated to
be $\pm$1.34 MeV/$c$.

\begin{figure}[ht]
\centering
\includegraphics[width=85mm]{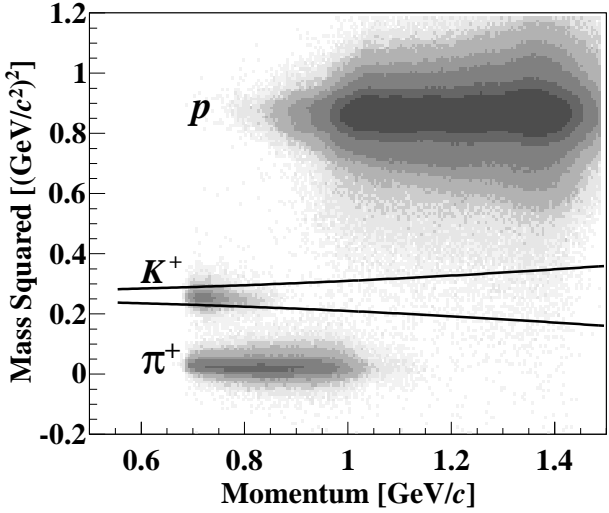}
\caption{Mass squared vs. momentum plot of the scattered particles measured by the SKS spectrometer in the $^6$Li($\pi^-,K^+$) reaction. Two curves in the figure show the momentum dependent 2$\sigma$ cut for the kaon selection.}
\label{mass2}
\end{figure}

%% H6L Production
We performed the measurement of the $^6$Li($\pi^-,K^+$) reaction at the
beam momentum of 1.2 GeV/$c$. We used high intensity beams of 
1.2--$1.4 \times 10^{7}$ pion/spill and the
effective total number of beam pions on the target was $1.4 \times 10^{12}$
taking into account the DAQ efficiency.
Since the $(\pi^-,K^+)$ reaction has no physical background and
the cross section of the reaction is very small, contaminations from
the miss-identification of $\pi^+$ and proton are the source of the backgrounds.
Figure \ref{mass2} shows the mass squared vs momentum plot
of the particles measured by the SKS spectrometer.
We selected kaons by a momentum dependent cut at $\pm 2\sigma$ of
the mass squared resolution as indicated with curves in the figure.
The contamination of protons in the $K^+$ cut region is at 1\% level
in the momentum range of 0.68--1.2 GeV/$c$, and the contamination of $\pi^+$ is 
negligibly small.

For quantitative discussions of the $^6$Li($\pi^-,K^+$) reaction, the double
differential cross section is derived from the following equation,
\begin{equation}
  \frac{d^2\sigma}{d\Omega d{\rm M}} = \frac{A}{N_A{\rho}x}\frac{n_K}{N_{\rm beam}\Delta\Omega{\Delta}{\rm M}\epsilon},
\end{equation}
where $n_K$ is the number of detected kaons in the missing-mass interval
${\Delta}{\rm M}$.
$N_A$ is the Avogadro number, and $A$ and ${\rho}x$ are
the atomic mass and the thickness in g/cm$^2$ of the target, respectively.
$N_{\rm beam}$ is the effective number of beam pions on the target including
the DAQ efficiency.
%where $A$, $N_A$ and ${\rho}x$ are the mass number, Avogadro number and 
%target thickness in g/cm$^2$, respectivly.
%$\epsilon$ is a total efficiency of experimental system.
%$N_{beam}$ and $N_K$ are the total beam pions on the target and the number
%of kaons detected, respectively.
%$n_K$ is the number of kaon within a mass resolution window, that is
%$N_K = \sum{n_k}$.
$\Delta\Omega$ is the angular acceptance of the SKS spectrometer.
%The SKS spectrometer acceptance, is estimated by a Monte Carlo
%simulation calculation based on {\sc geant4} package \cite{geant}, and
%is made the 2-dimensional map for the momentum and scattering angle.
%Typical acceptance around bound region is 100 msr.
An acceptance map of SKS in the 2-dimensional space of the momentum and the
emission angle of $K^+$ is estimated by a Monte Carlo simulation calculation
based on {\sc geant4} package \cite{geant}.
$\epsilon$ is the overall efficiency comes from detector and analysis
efficiencies estimated from experimental data.
%A typical SKS acceptance around the $\Lambda$ bound region, 0.9 GeV/$c$
%in the $K^+$ momentum, is 100 msr.
The differential cross section is also derived as follows,
\begin{equation}
  \frac{d\sigma}{d\Omega} = \frac{A}{N_A{\rho}x}\frac{N_K}{N_{\rm beam}\Delta\Omega \epsilon},
\end{equation}
where $N_K$ is the number of detected kaons, and $N_K = \sum n_K$ where the
summation runs over a spectral shape of signal events in the missing-mass
spectrum.

To confirm the validity of the procedure of the cross section calculation,
we estimated the cross section of the $p$($\pi^-,K^+$)$\Sigma^-$ reaction
with the same method.
\begin{figure}[ht]
\centering
\includegraphics[width=85mm]{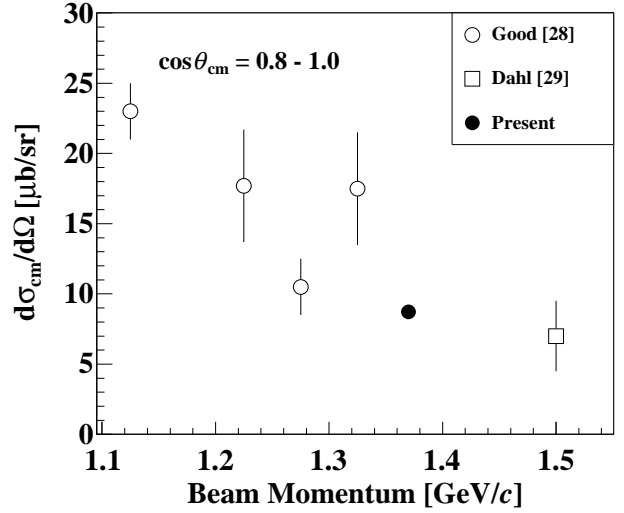}
\caption{
Differential cross section of the the $p(\pi^-,K^+)\Sigma^-$ reaction in the center of mass frame as a function of the beam momentum. The open circles and the open box are the cross sections reported by Good et al. \cite{good} and Dahl et al. \cite{dahl}, respectively. The full circle shows the present result.
}
\label{CS_SigmaN}
\end{figure}
Figure \ref{CS_SigmaN} shows the estimated differential cross section,
$d\sigma_{\rm cm}$/$d\Omega$, in the center of mass frame (full circle)
in the angular range of $\cos\theta_{\rm cm}$=0.8--1.0 together with
the cross sections reported by Good et al. \cite{good} (open circles)
and Dahl et al. \cite{dahl} (open box).
The differential cross sections gradually decrease with the increase
of the beam momentum, and the
present result is consistent with the general trend.

\begin{figure}[ht]
  \centering
  \includegraphics[width=85mm]{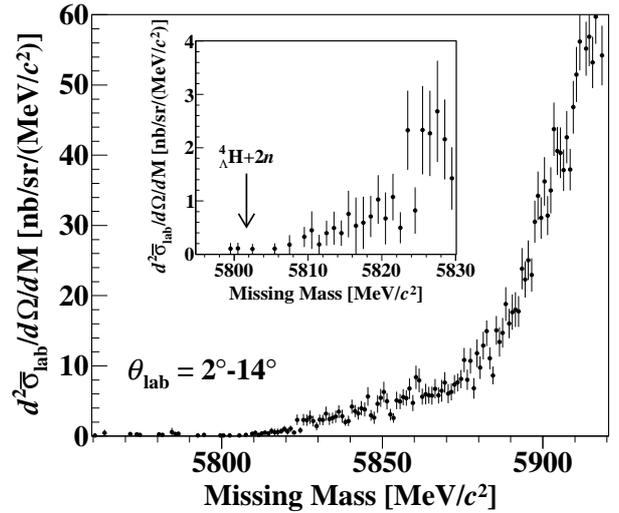}
  \caption{Missing-mass spectrum of the $^6$Li$(\pi^-,K^+)$ reaction at 1.2 GeV/$c$. The ordinate shows the double differential cross section averaged over the angular range from 2 to 14 degrees. A magnified view around the $\Lambda$ bound region is shown in the inset. The arrow labeled as $^4_{\Lambda}$H$+2n$ shows the particle decay threshold (5801.7 MeV/$c^2$).}
  \label{CS6HL}
\end{figure}

Figure \ref{CS6HL} shows the missing-mass spectrum of
the $^6$Li$(\pi^-,K^+)$ reaction.
The vertical axis shows the double differential cross section
in the laboratory frame averaged over the scattering angle
from 2$^{\circ}$ to 14$^{\circ}$,
$d^{2}\overline{\sigma}_{\rm lab}$/$d\Omega/d$M in a unit of nb/sr/(MeV/$c^2$).
%The anglular range is selected due to relatively small ambiguity in the
%acceptance estimation of the SKS spectrometer.
The estimation of the spectrometer acceptance has small ambiguity in the
selected angular range.
The uncertainty of the missing-mass scale 
is $\pm$1.26 MeV/$c^2$ which is estimated from the beam momentum uncertainty
$\pm$1.34 MeV/$c$.
The continuum of the unbound $\Lambda$ formation reaction
and the component of the $\Sigma^-$ quasi-free production reaction are
observed in the missing-mass regions of 5810--5880 MeV/$c^2$ and
above 5880 MeV/$c^2$, respectively.
A magnified view in the missing-mass range of 5795--5830 MeV/$c^2$
is shown in the inset.
Around the  $^4_{\Lambda}$H$+2n$ particle decay
threshold indicated by the arrow (5801.7 MeV/$c^2$),
no significant peak structure is observed.

%As shown in Fig.\ref{CS6HL}, there is a finite cross section in
%the missing-mass region around the $^4_{\Lambda}$H$+2n$ threshold.
%Since the mass resolution is 3.2 MeV/$c^2$, if it is considered the resolution,
%the corresponding number of event is 3 events around the region of the
%$^4_{\Lambda}$H$+2n$ threshold.
As the first step of the calculation of an upper limit
of the differential cross section,
$d\overline{\sigma}_{\rm lab}/d\Omega$, for a state of $^6_{\Lambda}$H,
we estimated the number of observed events in the missing-mass region around the
$^4_{\Lambda}$H$+2n$ threshold.
For the estimation of the number of events which associate to the production
of a state, we set a missing-mass window of $\pm2 \sigma$, where $\sigma$ is
rms of the missing-mass resolution ($\sigma=1.36$ MeV/$c^2$).
There are 3 events in the threshold region within the missing-mass window. 
Therefore, we interpret the 3 events as observed events which
include background and possible signal events.

Another necessary information for the upper limit estimation is the number
of background events. 
Population of events are observed in the missing-mass region lower than
5780 MeV/$c^2$ where we do not expect any physical backgrounds.
The level of the event population averaged in the range from 5700 to 5780
MeV/$c^2$ is $(0.39\pm0.05)$ events per 1 MeV/$c^2$.
These events are instrumental background due to particle
miss-identifications, and a similar level of backgrounds are expected
also at the $^4_{\Lambda}$H+2$n$ threshold.
The number of background events is estimated by multiplying the
background level and the width of the missing-mass window.

If the Poisson statistics is used, the upper limit of the number of signal
events is estimated to be 4.80 at 90\% confidence level.
Although we assume a flat distribution of the background events,
the missing-mass dependence of the background level is not well known due to
the low statistics.
Therefore, we employ another upper limit of 6.68 events coming from the
background free hypothesis as a conservative estimation.

As shown in Fig.\ref{CS6HL}, the differential cross sections are
roughly 0.1 nb/sr per event for the observed 3 events.
However, the differential
cross sections largely depend on the observed scattering angles and may have
a statistical bias.
If an event happens to be observed at the forward angle,
the cross section is underestimated, and vice versa, in our setup.
To avoid the bias of the observation, we estimate the differential
cross section averaged over the selected angular range, from 2$^\circ$ to
14$^\circ$, and we obtain a value of 0.18 nb/sr for 1 event.
By using the value, the upper limit of the differential cross section averaged
in the scattering angle from 2$^\circ$ to 14$^\circ$ is estimated to be 1.2
nb/sr at 90\% confidence level.

\section{Discussion}
In our measurement, neither significant peak structure nor a large yield is
observed around the $^4_{\Lambda}$H$+2n$ particle decay threshold
in the missing-mass spectrum of the $^6$Li($\pi^-,K^+$) reaction.
The $^6_{\Lambda}$H hypernucleus is believed to have the
$^4_{\Lambda}$H$+2n$ structure dominantly and to have the 0$^+$ ground and
the 1$^+$ excited states which are analogous to the 0$^+$ and
1$^+$ spin-doublets in $^4_{\Lambda}$H.
As far as the FINUDA result is concerned \cite{finuda2},
the observed $^6_{\Lambda}$H candidate events were
interpreted as the primary population of the excited 1$^+$ state
by the ($K^-_{stopped},\pi^+$) reaction followed by the $\gamma$-ray
transition to the ground 0$^+$ state because the direct population of the 0$^+$
state should be suppressed due to the small spin-flip
amplitude.
The FINUDA results indicate that the excited 1$^+$ state,
whose excitation energy is estimated to be 1 MeV,
should be particle bound, otherwise the
$\gamma$-ray transition to the $0^+$ ground state should be impossible.
If the 1$^+$ state is bound and the production cross section is comparable with
that for $^{10}_{~\Lambda}$Li of 11.3 nb/sr, more than 60 events should be
observed as a peak in the $\Lambda$ bound region in the missing-mass spectrum of
the $^6$Li($\pi^-,K^+$) reaction.
Therefore, our observation is in conflict with the simple interpretation
of the FINUDA observation.

Hiyama et al. \cite{hiyama} suggested that the both 0$^+$ and 1$^+$ states
are unbound. If it is the case, production 
cross sections may be smaller than the sensitivity of
our measurement due to the broad wave-functions of the unbound states
which have small overlaps with the wave-function in the initial $^6$Li nucleus.
Gal and Millener \cite{gal} discussed another possible interpretation of
the FINUDA observation.
They suggested a possibility of an unbound 1$^+$ state, whose particle decay 
width is extremely small and comparable with that of the $M1$ $\gamma$
decay to the ground 0$^+$ state due to kinematical
and dynamical suppression of the emission of two neutrons from the
1$^+$ excited state.
The FINUDA collaboration also discussed another scenario for the spectrum
of $^6_{\Lambda}$H in which two out of the three candidate events came from the
population of the spin-triplet states, $1^+$, $2^+$ and  $3^+$, at around
3 MeV excitation \cite{finuda3}.
Therefore, it is interesting to compare our upper limit,
1.2 nb/sr, with quantitative
theoretical estimations of the production cross sections because the
cross sections are sensitive to the binding energies and
the wave-functions of the low-lying states.

\section{Summary}
We performed the measurement of the $^6$Li($\pi^-,K^+$) reaction at the
beam momentum of 1.2 GeV/$c$.
No significant peak structure is observed around the $^4_{\Lambda}$H$+2n$
threshold.
The upper limit of the differential cross section in the scattering angle
from 2$^\circ$ to 14$^\circ$ is estimated to be 1.2 nb/sr at the 90\% confidence
level.
The result does not favor the simple interpretation of the FINUDA
observation and it suggests reconsideration of the structure of the
$^6_{\Lambda}$H hypernucleus.
More clear insight of the $^6_{\Lambda}$H structure may be obtained
by comparing our cross section upper limit with quantitative theoretical
calculations.

\section*{Acknowledgements}

We would like to acknowledge the staff of the Hadron beam line
and J-PARC accelerator groups for their efforts of improving the beam
quality and keeping the stable operation during the beam time.
We also thank Professors  Y. Akaishi, T. Harada, A. Gal and E. Hiyama,
for the fruitful discussions. 
The experiment was supported by the Grant-in-Aid
for Scientific Research (KAKENHI),
for Scientific Research on Innovative Areas No.~24105003, 
for Young Scientist (A) No.~23684011,
for Basic Research (Young Researcher) No.~2010-0004752 from National
Research Foundation in Korea
and for Scientific Research (C) No.~24540305 from the Ministry of Education, Culture, Sports,
Science and Technology in Japan.
We acknowledge support from WCU program, National Research Foundation,
Center for Korean J-PARC Users, and the Ministry of Education,
Science, and Technology (Korea).



%% References


%\bibliographystyle{elsarticle-num}



\end{document}